\documentclass[conference]{IEEEtran}
\IEEEoverridecommandlockouts

\newif\ifEditMode
\EditModetrue

\usepackage{algorithm2e}
\RestyleAlgo{ruled}

\usepackage{cite}
\usepackage{amsmath,amssymb,amsfonts}
\usepackage{graphicx}
\usepackage{textcomp}
\usepackage{aliases}
\usepackage{preamble}

\usepackage{amsthm}

\usepackage[table]{xcolor}
\usepackage{colortbl}
\usepackage{booktabs}

\def\BibTeX{{\rm B\kern-.05em{\sc i\kern-.025em b}\kern-.08em
    T\kern-.1667em\lower.7ex\hbox{E}\kern-.125emX}}
\begin{document}

\title{On the Centralization of Governance Power
in Decentralized Autonomous Organizations}

\author{\IEEEauthorblockN{Vabuk Pahari}
\IEEEauthorblockA{
\textit{Max Planck Institute for Software Systems}\\
Germany \\
vpahari@mpi-sws.org}
\and
\IEEEauthorblockN{Balakrishnan Chandrasekaran}
\IEEEauthorblockA{
\textit{Vrije Universiteit Amsterdam}\\
Netherlands \\
b.chandrasekaran@vu.nl}
\and
\IEEEauthorblockN{Johnnatan Messias}
\IEEEauthorblockA{
\textit{Max Planck Institute for Software Systems}\\
Germany \\
johnme@mpi-sws.org}
\and
\IEEEauthorblockN{Krishna P. Gummadi}
\IEEEauthorblockA{
\textit{Max Planck Institute for Software Systems}\\
Germany \\
gummadi@mpi-sws.org}
\and
\IEEEauthorblockN{Abhisek Dash}
\IEEEauthorblockA{
\textit{Max Planck Institute for Software Systems}\\
Germany \\
adash@mpi-sws.org}
}

\maketitle

\begin{abstract}

A decentralized autonomous organization (DAO) is a governing entity that empowers its stakeholders (i.e., users who hold one or more of its tokens) to manage blockchain-based protocols (i.e., smart contracts) collaboratively.
The governance of a DAO is explicitly encoded in the DAO's \emph{governance contract}, which defines how stakeholders participate in governance and how much influence (or voting power) they have in any decision.
While decentralization and autonomy are the fundamental tenets of a DAO's design, empirical evidence suggests that in practice governance is often highly centralized.
In this work, we study the designs and implementations of \surveysz{} public and actively used DAOs, with substantially large capital, deployed on Ethereum.
We identify how three key governance mechanisms---token registration, staking, and delegation---originally introduced to improve security or participation, contribute to the concentration of voting power. 
Unlike prior work on centralization of voting power in specific DAOs, our findings reveal that these governance mechanisms of DAOs themselves systematically reinforce centralization.
By elucidating the relationship between governance design and voting centralization, this work advances the understanding of DAO governance structures and highlights the inherent trade-offs between decentralization, security, and usability of DAOs.

\end{abstract}
\section{Introduction}\label{s:introduction}

Decentralized Finance (DeFi) promises to revolutionize traditional finance by
disintermediating crucial services such as banking, derivatives, and
insurance~\cite{adams2021uniswap,Daian@S&P20,Qin@FC21,Perez@FC21}.
Instead of relying on any trusted party, formally specified protocols mediate the interactions between users with no \stress{a priori} trust relationships.
These protocols are written in transparent smart contracts, which are deployed on public and permission-less blockchains, allowing any user to audit the code.
DeFi protocols have quickly gained widespread adoption (with millions of users) and attracted substantial investments (with more than 100 billion USD locked in various DeFi smart contracts) since their inception in 2017~\cite{defillama}.
Their potential for creating a trustworthy and transparent financial ecosystem, however, crucially rests on how the underlying smart contracts are \stress{managed}.

\paraib{Governance Contracts} 
Managing a smart contract entails two key tasks:
(i) maintenance of the contract, which is concerned with how the contract is \stress{updated} (e.g., to adapt to market conditions or fix bugs) and \stress{changed} (e.g., to offer new financial instruments) over time, and
(ii) financing of the contract's operations, which is concerned with tracking the profits and losses accumulated in the protocol's treasury as well as compensating the people working on maintaining the contract.
These management tasks are transparently specified in a \stress{governance contract}, which formally defines \stress{(i) who has the power to make the management decisions} and \stress{(ii) how the decision makers exercise their power}.

\paraib{Decentralized Governance in Theory}
A governance contract vests the power to manage a smart contract, typically, in the holders of governance tokens.
A \stress{decentralized autonomous organization (DAO)} is simply the abstract entity comprising all holders of the governance tokens.
While recent legislation in a few places, e.g., Wyoming, Vermont and Cayman Islands, allow a DAO to be registered as a legal entity~\cite{wyoming-dao,cayman-islands-dao,vermont-law}, there exists no widely accepted formal or legal definition for what constitutes a DAO.
Yet, governance contracts underlying DAOs share two characteristic features:
(i) \stress{autonomy of governance}, or \stress{self-governance}, whereby the current governance token holders have the power to amend the governance contract themselves and
(ii) \stress{decentralization of governance}, whereby all decisions, are arrived at through a referendum-style vote by all governance token holders.
This decentralized decision making crucially distinguishes DAOs from traditional organizations, which recruit managers and entrust them with making such decisions.

\paraib{Decentralized Governance in Practice}
Recent studies \cite{messias2023gov, fritsch@2022votingpower, sharma2023unpacking} have shown that in many DAOs only a small number (and fraction) of all governance token holders participate in voting on governance proposals.
Messias \ea \cite{messias2023gov} found, for instance, that in Compound, on average, as few as 3 voters controlled the majority of exercised votes and, hence, control the outcome of a proposal. 
Similarly, Fritsch \ea \cite{fritsch@2022votingpower} observed that in most DAOs, just three voting wallets held more than 50\% of exercised voting power.
Low stakeholder participation risks concentrating the decision making in hands of a few, undermining the ideals of decentralized governance. 
Prior work have speculated about the causes of this centralization---attributing it to factors such as the high cost of voting, a small number of voters with high voting power, voter collusion, or voter apathy. 
Much less attention has been paid, however, to the effects of mechanisms adopted by DAOs to protect the governance voting from undue interference and malicious attacks, while encouraging eligible voter participation.
In this work, we focus on three widely used mechanisms namely, {\it voter registration, token staking, and proxy voting} that allow DAOs to regulate which of their token holders can exercise their right to vote and how.
These mechanisms have been adopted because in practice (a) not all governance token holders should have the right to vote, (b) the interests of the token holders who are willing to stake their tokens (i.e., pledge to not sell them until some future time) are better aligned with the long-term interests of the DAOs, and (c) given the frequency of votes, voter participation may improve if they are allowed to delegate their vote and vote via proxies.
The central finding of this work is that while these mechanisms are well-motivated, in practice, they have many complex unintended consequences, which include both the disenfranchisement of a number of legitimate token holders and the concentration the governance power in the hands of a few token holders.
Our analysis shows that how the design of these mechanisms themselves leads to increased centralization, and how they can be altered to allow greater decentralization.

In this work, we investigated the governance contracts underlying \surveysz{} DAOs with the largest treasury sizes~\cite{deepdao}.
%
We demonstrate that governance mechanisms introduced to enhance the security and usability of DAOs---token registration, staking, and delegation---can inherently lead to the centralization of voting power in practice. 
We summarize our key contributions as follows.

\point{}
We review the (software) implementations of the smart contracts underlying
\surveysz{} DAOs and peruse the developer documentations to identify the key
design choices used in DAOs. 

\point{}
We analyze the distribution and structure of token ownership in DAOs (\S\ref{s:exercise-governance-power}), drawing parallels between governance tokens and different classes of shares used in traditional corporate governance. 
Our findings clarify current practices concerning the treatment of governance tokens and suggest how they should be managed with respect to their governance and monetary rights.

\point{}
We investigate the different mechanisms used by DAOs in granting governance and monetary rights (\S\ref{s:user-parameters}-\ref{s:voting-power-distribution}), and how these choices contribute to the centralization of voting power. Where possible, we suggest ways to alter the mechanisms to enhance decentralization of voting power.

\point{}
We will release the data gathered and the code for analyzing them as open-source artifacts upon publication.

\section{Methodology}\label{s:methodology}

In this study, we selected \surveysz{} DAOs based on four criterias.

\point{}
We focused on DAOs that have their relevant governance contracts deployed on Ethereum, the largest blockchain in terms of Total Value Locked (TVL) on-chain \cite{defillama}.

\point{}
We selected DAOs with the largest treasury sizes, as reported by DeepDAO~\cite{deepdao} in May 2024. 
Many DAOs correspond to the leading DeFi projects in terms of Total Value Locked (TVL), reflecting strong user trust and a significant impact on the DeFi landscape~\cite{deepdao}.

\point{}
We focused on \stress{public} DAOs, where \stress{anybody} can participate in governance by acquiring governance tokens, either by buying it on a market or by doing some work for the DAO.
We excluded, hence, DAOs whose governance structures were \stress{private}, e.g., Graph Protocol~\cite{graph@get-started}.
This form of governance, where all decisions are made by only a small subset of users, and other users have neither the voting rights nor any role in governance is antithetical to the notion of decentralization.

\point{}
We excluded DAOs that have not had a vote in the last $6$ months (as of May 2024), essentially excluding inactive projects, with likely `stale' (i.e., not kept up-to-date) documentations on their policies or removed by law. 

\stress{Essentially, we selected DAOs that were public, active, held substantially large capital, widely used and trusted by users, and had enormous potential to be coupled or integrated with other emerging DeFi applications.}

\paraib {Code and Data Analyzed}
For each of the \surveysz{} DAOs selected, we parsed the official documentations to fetch contract addresses and their off-chain voting platforms.
Then, we gathered all their governance-related contracts such as the governance contracts, governance token contracts, and staking contracts using Etherscan.
DAOs publish the relevant contract code on Etherscan, and Etherscan validates the correctness of the code by compiling it and checking that the bytecode of the submitted code matches the bytecode deployed on Ethereum.
We used Etherscan to identify DAO's contract addresses, associated treasury and vesting addresses. 
We also gather all addresses identified as a CEX wallet address from 15 Centralized Exchanges (CEX) from their websites and Etherscan, namely: Coinbase, Binance, ByBit, OKX, Upbit, Kraken, Gate.io, Bitfinex, KuCoin, Crypto.com, MEXC, Gemini, Huboi, BitGet and Robinhood.
We identified all pools associated with Decentralized Exchanges (DEX)---Uniswap V1, V2 and V3, Sushiswap V1 and V2, and Balancer---and Lendinng Protocols (Compound and AAVE).
We ran an Ethereum Archive node (Erigon Client) to gather data for our empirical analysis from these contracts at block number \num{23270000} (September 1st, 2025).
Lastly, we collected off-chain voting data of DAOs using the official Snapshot API~\cite{Snapshot,Snapshot-api}.

\section{Ownership Structures: Corporations vs. DAOs}\label{s:exercise-governance-power}

\begin{table*}
\tabcap{Table depicting the number of Tokens Authorized, Issued, in Treasury, Allocated, Vested, Outstanding and Registered. Percentages are calculated based on the Authorized Tokens. DAOs in color are DAOs where new governance tokens can be minted: \textbf{\color{red}Red} means there is \textbf{no} limit on the number of new tokens, \textbf{\color{green}Green} means there is a limit on the total number of new tokens, 
and \textbf{\color{blue} Blue} means that there is a cap on the total number of new tokens in a year (but no caps overall). The rest of the DAOs cannot create new tokens. 
(-) means that the DAO has no registration or no staking.
}
\label{tab:table-registered-tokens}
\begin{center}
\resizebox{\textwidth}{!}{
\begin{tabular}{c l r r r r r r r r r}
\toprule
Id & DAO & Authorized & Issued & Treasury (\%)  & Unvested (\%) & Outstanding (\%) & Registered (\%) & Staking (\%) & Floating (\%)  \\ [0.5ex] 
\midrule
\arrayrulecolor{lightgray}

1 & \color{blue}uniswap & $\infty$ & \num{1000000000.0} & 37.126 & 0.0 & 62.874 & 19.209 & (--) & 62.874 \\
\hline
2 & \color{blue}ens & $\infty$ & \num{100000000.0} & 14.291 & 48.943 & 36.765 & 4.076 & (--) & 36.765 \\
\hline
3 & \color{red}maker & $\infty$ & \num{392795.9} & 0.0 & 0.0 & 100.0 & 1.533 & (--) & 100.0 \\
\hline
4 & \color{red}lido & $\infty$ & \num{1000000000.0} & 10.423 & 18.842 & 70.735 & (--) & (--) & 70.735 \\
\hline
5 & \color{green}frax & \num{100000000} & \num{99681495.6} & 8.764 & 0.0 & 91.236 & 24.96 & 24.96 & 66.276 \\
\hline
6 & aave & \num{16000000.0} & \num{16000000.0} & 2.954 & 0.0 & 97.046 & (--) & 17.629 & 79.417 \\
\hline
7 & compound & \num{10000000.0} & \num{10000000.0} & 5.193 & 0.0 & 94.807 & 19.208 & (--) & 94.807 \\
\hline
8 & radicle & \num{99998580.0} & \num{99998580.0} & 48.423 & 3.274 & 48.302 & 10.613 & (--) & 48.302 \\
\hline
9 & zrx & \num{1000000000.0} & \num{1000000000.0} & 10.441 & 5.325 & 84.233 & 5.171 & 5.171 & 79.062 \\
\hline
10 & \color{blue}gitcoin & $\infty$ & \num{100000000.0} & 3.615 & 0.0 & 96.385 & 8.41 & (--) & 96.385 \\
\hline
11 & \color{green}silo & \num{1000000000} & \num{174885739.1} & 0.0 & 52.387 & 47.613 & 14.068 & (--) & 47.613 \\
\hline
12 & lyra & \num{1000000000.0} & \num{1000000000.0} & 38.979 & 10.859 & 50.161 & 25.764 & 25.764 & 24.398 \\
\hline
13 & \color{red}api3 & $\infty$ & \num{151252891.6} & 7.659 & 0.504 & 91.837 & 39.946 & 39.946 & 51.891 \\
\hline
14 & \color{blue}ampleforth & $\infty$ & \num{15297897.1} & 12.104 & 0.477 & 87.42 & 5.371 & (--) & 87.42 \\
\hline
15 & \color{blue}instadapp & $\infty$ & \num{100000000.0} & 23.247 & 2.973 & 73.781 & 17.685 & (--) & 73.781 \\
\hline
16 & \color{red}rari & $\infty$ & \num{25000000.0} & 14.513 & 0.0 & 85.487 & 0.839 & 0.839 & 84.648 \\
\hline
17 & \color{red}nouns & $\infty$ & \num{1634.0} & 32.864 & 0.0 & 67.136 & 100.0 & (--) & 67.136 \\
\hline
18 & \color{green}curve & \num{3030303031} & \num{2300143704.3} & 1.596 & 1.075 & 97.329 & 37.745 & 37.745 & 59.584 \\
\hline
19 & \color{red}origin & $\infty$ & \num{1409664846.0} & 10.594 & 0.0 & 89.406 & 37.485 & 37.485 & 51.921 \\
\hline
20 & \color{red}hop & $\infty$ & \num{1000000000.0} & 51.154 & 26.985 & 21.861 & 2.369 & (--) & 21.861 \\
\hline
21 & \color{blue}cryptex & $\infty$ & \num{10000000.0} & 24.179 & 0.25 & 75.571 & 16.368 & (--) & 75.571 \\
\hline
22 & \color{blue}angle & $\infty$ & \num{1000000000.0} & 35.056 & 20.624 & 44.32 & 1.621 & 1.621 & 42.699 \\
\hline
23 & \color{red}dxdao & $\infty$ & \num{2530180.8} & 0.0 & 0.0 & 100.0 & 100.0 & (--) & 100.0 \\
\hline
24 & \color{red}nexus & $\infty$ & \num{2204746.5} & 1.524 & 0.0 & 98.476 & 72.282 & (--) & 98.476 \\
\hline
25 & \color{red}goldfinch & $\infty$ & \num{114285714.0} & 17.937 & 0.0 & 82.063 & 0.43 & (--) & 82.063 \\
\hline
26 & \color{red}paragonsdao & $\infty$ & \num{143227431.9} & 0.0 & 1.377 & 98.623 & (--) & 1.687 & 96.936 \\
\hline
27 & \color{red}illuvium & $\infty$ & \num{9072238.3} & 0.295 & 30.135 & 69.57 & 10.567 & 7.911 & 61.659 \\
\hline
28 & \color{red}superrare & $\infty$ & \num{1000000000.0} & 17.564 & 0.0 & 82.436 & (--) & (--) & 82.436 \\
\hline
29 & \color{blue}mantle & $\infty$ & \num{6219316795.0} & 47.693 & 0.0 & 52.307 & 5.13 & (--) & 52.307 \\
\hline
30 & reasearchhub & \num{1000000000.0} & \num{1000000000.0} & 79.524 & 0.276 & 20.2 & (--) & (--) & 20.2 \\
\hline
31 & stargate & \num{1000000000.0} & \num{1000000000.0} & 32.795 & 0.0 & 67.205 & 1.075 & 1.075 & 66.13 \\
\hline
32 & \color{red}uma & $\infty$ & \num{125950840.9} & 27.323 & 0.0 & 72.677 & 26.527 & 26.527 & 46.15 \\
\hline
33 & \color{blue}cow & $\infty$ & \num{1000000000.0} & 40.942 & 20.985 & 38.073 & (--) & (--) & 38.073 \\
\hline
34 & \color{red}sturdy & $\infty$ & \num{100000000.0} & 79.596 & 0.0 & 20.404 & (--) & (--) & 20.404 \\
\hline
35 & \color{blue}euler & $\infty$ & \num{27182818.3} & 24.517 & 2.701 & 72.782 & 9.025 & (--) & 72.782 \\
\hline
36 & safe & \num{1000000000.0} & \num{1000000000.0} & 14.775 & 52.327 & 32.898 & (--) & 7.039 & 25.86 \\
\hline
37 & \color{green}tokenlon & \num{200000000} & \num{140451028.8} & 6.72 & 0.327 & 92.953 & (--) & 61.039 & 31.914 \\
\hline
38 & botto & \num{100000000.0} & \num{100000000.0} & 13.391 & 1.72 & 84.889 & 34.391 & (--) & 84.889 \\
\hline
39 & \color{green}balancer & \num{95634448} & \num{69651658.5} & 6.633 & 0.0 & 93.367 & 31.756 & 31.756 & 61.611 \\
\hline
40 & \color{red}sushi & $\infty$ & \num{285195141.8} & 4.827 & 0.0 & 95.173 & 5.402 & (--) & 95.173 \\
\hline
41 & gearbox & \num{10000000000.0} & \num{10000000000.0} & 33.873 & 15.967 & 50.16 & 7.037 & 2.902 & 47.258 \\
\hline
42 & \color{blue}paraswap & $\infty$ & \num{2000000000.0} & 8.268 & 20.257 & 71.475 & 22.529 & 4.221 & 67.254 \\
\hline
43 & \color{red}alchemix & $\infty$ & \num{3079181.6} & 4.441 & 1.062 & 94.497 & (--) & (--) & 94.497 \\
\hline
44 & oneinch & \num{1500000000.0} & \num{1500000000.0} & 6.376 & 3.862 & 89.763 & 17.83 & 17.83 & 71.933 \\
\hline
45 & shutter & \num{1000000000.0} & \num{1000000000.0} & 56.074 & 22.715 & 21.211 & 23.077 & (--) & 21.211 \\
\hline
46 & \color{red}yearn & $\infty$ & \num{36666.0} & 7.683 & 0.0 & 92.317 & 4.311 & 4.311 & 88.005 \\
\hline
47 & shapeshift & \num{1000001337.0} & \num{1000001337.0} & 17.785 & 7.337 & 74.878 & (--) & (--) & 74.878 \\
\hline
48 & \color{red}decentraland & $\infty$ & \num{2193179327.3} & 2.305 & 10.637 & 87.059 & (--) & (--) & 87.059 \\
\hline

\arrayrulecolor{black}
\bottomrule
\end{tabular}
}
\end{center}
\end{table*}

To develop a better understanding of the ownership structures and governance mechanisms in DAOs, we propose to compare and contrast them with those of corporations.
Public corporations, which are entities created and recognized by law and owned and governed by shareholders, have existed and evolved over the course of the past four centuries. 
Ideally, DAOs should have efficient and effective mechanisms to enable the ownership and governance structures of today's public corporations and potentially, offer support for more richer structures. 
As our discussion below shows, it can be quite challenging to support existing basic corporate ownership and governance structures via DAOs.

\subsection{Types of corporate shares} 
In corporations, ownership is recognized via a person holding a share. 
Typically, the holder of a share has both governance and monetary rights, i.e., the right to vote on decisions affecting the corporation, and the right to monetize the share as an asset (by receiving a portion of the dividends distributed, or by pledging the shares as collateral with lenders, or by selling the shares on markets). 
However, in corporate governance, not all shares are created equal -- they can take different forms and can have distinct rights associated with them.
Below, we distinguish between different types of corporate shares.

{\bf Authorized shares} are the maximum number of shares a company is legally allowed to issue, as specified in its articles of incorporation. 
They create a legal ceiling and establish a lower bound on the fraction of ownership represented by a share in the case when all of them are issued.
However, typically, only a fraction of authorized shares are issued and in circulation at any given time\cite{authorized-outstanding-shares@Investopedia}.

{\bf Issued shares} are the portion of authorized shares that the company has actually distributed to shareholders. 
These can be sold to investors during an IPO, or private placement, or other fundraising activities. 
They can also be offered to employees in lieu of their compensation with restrictions, i.e., conditioned on the employees meeting certain goals within sometime.
While issued shares represent ownership in the company, the following two types of issued shares carry restricted rights. 

{\bf Treasury shares} are issued shares that are held by the corporate treasury account.
These could be issued shares that remained unsold to investors or repurchased shares to reduce circulation \cite{treasury-shares@Investopedia}. 
As they are controlled by the corporation itself, they are held without voting or dividend rights.

{\bf Unvested shares} are issued (usually to employees) but are subject to a vesting schedule---the employee must meet certain conditions before gaining full ownership. 
Until vested, these shares often have restrictions on use and transferability.
Hence, they too are held without voting and dividend rights.

{\bf Outstanding shares} are held by shareholders and form the basis for ownership, governance power and dividends \cite{authorized-outstanding-shares@Investopedia,issued-outstanding-shares@Investopedia}.
Some corporations issue shares that belong to different classes, with shares in {\bf preferred} classes having more or less voting and/or monetary rights than shares in the {\bf common} class \cite{common-shares@Investopedia}.   
Not all outstanding shares are freely tradeable in public markets and the term 
{\bf floating} shares refers to outstanding shares that are freely tradable in public markets \cite{floating-shares@Investopedia}.

\subsection{Types of DAO tokens}\label{ss:token-types}
In DAOs, ownership is represented by a wallet holding a governance token.
Typically, governance tokens carry with them both governance and monetary rights.
However, like corporate shares, not all DAO tokens should bestow both these rights to their holders.
Unfortunately, as all governance tokens are fundamentally fungible (i.e., interchangeable and indistinguishable), DAOs need additional mechanisms to distinguish between holders of different types of tokens, i.e., authorized, issued, treasury, unvested, and outstanding, preferred, and floating tokens. 
We discuss the challenges below.

{\bf Authorized \& Issued tokens:} ~\autoref{tab:table-registered-tokens} shows the number of authorized and issued tokens for the 48 DAOs in our study.  
We obtained these numbers by examining the governance token contract of the DAOs and the functions that lead to the creation (i.e., minting and issuance) of their tokens. 
Of the 48 DAOs, only 18 enforce a cap on total number of authorized tokens, while 30 impose no limit on the total number of tokens that can be created.
Of the 18 DAOs limiting authorized tokens, 13 have already issued (created) all of their authorized tokens.
Of the 30 DAOs allowing unlimited authorized tokens, 11 impose a limit on the annual growth in the supply of authorized tokens (typically 1–2\% of the current supply), while the remaining 19 DAOs allow arbitrary dilution of the ownership represented by a token (which raises potential governance concerns).

{\bf Treasury and Unvested tokens:} 
As many DAOs are still in their infancy, they have a significant stock of issued tokens both in their treasury and vesting contracts.
As neither treasury nor unvested tokens should carry governance or monetary rights, it is important to (a) have the ability to reliably identify all treasury wallets and vesting contracts of a DAO and (b) have mechanisms to block the tokens from participating in governance vote or receiving dividends. 
However, we lack reliable ways to identify and restrict all treasury and unvested token holding wallets.

Hence, we relied on heuristics to identify treasury and vesting contract wallets to estimate a lower-bound on fraction of issued shares held by them. 
To identify a DAO’s treasury wallets, we examined the top 50 token holders and selected addresses labeled as “treasury” on Etherscan \cite{Etherscan@ETH-explorer} or controlled by the DAO’s governance contract.
For vesting contracts, we also reviewed the top 50 holders to find contracts with names indicating vesting or locking and verified that tokens could not be removed.
To expand this list, we used Etherscan's ``Similar Contracts Search'' \cite{etherscan-similar} to locate other contracts with identical bytecode, identifying additional vesting contracts associated to the DAO.
~\autoref{tab:table-registered-tokens} shows the estimates of DAO treasury and unvested tokens obtained using our heuristics. We find that 28 (of the 48) DAOs have significant fraction (at least 10\% or more) of their issued tokens in their treasury wallets, while 27 DAOs have tokens locked in vesting contracts for some time.

To ensure that treasury and unvested token holding wallets do not vote on governance issues, many DAOs require their {\it wallets to register to vote}. 
Tokens can then be blocked from voting by having their wallets controlled by smart contracts that do not offer the wallet registration functionality. 
Although these tokens should have no governance or monetary rights, we find that treasury tokens of 1 DAO (\#2)\cite{ens-veto-forum} and unvested tokens in 5 DAOs (\#15,33,35,41,45) have been allowed to register to participate in governance votes, which raises serious concerns about their governance practices.

{\bf Outstanding and Floating tokens:} 
Outstanding tokens consist of issued tokens that are not treasury or unvested. 
These tokens carry both governance and monetary rights, which can be exercised when their wallets register to vote.
~\autoref{tab:table-registered-tokens} shows that across many DAOs, only a small fraction of all outstanding tokens are registered and, hence, are eligible to vote.
Less than 50\% of eligible tokens have registered to vote in 32 out of 36 DAOs requiring registration, raising concerns about concentration of voting power. 
We will investigate why voter registration mechanisms today disenfranchise many legitimate outstanding token holders in \autoref{s:user-parameters}.

Furthermore, in 15 DAOs, 
voter registration requires the tokens to be {\it staked}, i.e., locked for a pre-specified period of time, when they cannot be traded at will. 
The token holders of these DAOs have to choose between exercising their governance and monetary rights--to obtain governance rights, they must be willing to forgo their monetary right to sell during the staking period and vice-versa. 
We will analyze these trade-offs in depth later in \autoref{s:voting-power}.
The choices presented by staking raise concerns about different types of centralization of governance and monetary power-- if token holders choose not to stake, they risk concentrating governance power in fewer holders (the median fraction of staked tokens in the 15 DAOs is 27.4\%. 
If token holders choose to stake, the staked tokens are not freely tradable during this lock-in period, sharply reducing the fraction of \stress{floating} tokens (the median floating supply 66.8\% or one-third less than authorized/issued tokens), leaving their liquidity and price stability to decisions by fewer holders. 

Finally, estimating the size of outstanding tokens matters for accurately assessing a DAO’s market capitalization. 
Current tracking platforms often rely on authorized or issued token counts, which can significantly inflate valuations --- particularly in young DAOs where most tokens remain in treasury \cite{saggese2025verifiabilitytotalvaluelocked,outstanding-supply@artemis}.
~\autoref{tab:table-registered-tokens} shows that the median outstanding supply (excluding treasury and unvested tokens) is only 78.9\% of authorized tokens, implying that current estimates of market capitalization of DAOs based on authorized tokens may be overstating by at least one-fifth.

\noindent {\bf Summary:} By comparing and contrasting ownership structures and governance mechanisms of DAOs with those of traditional corporations, our discussion surfaces multiple concerns that have been previously overlooked in the literature: (i) many DAOs lack reliable mechanisms to block treasury and unvested token holders from participating in governance votes. (ii) existing mechanisms, particularly voter registration and token staking, result in a limited and small fraction of token holders having the eligibility to vote. We will investigate the mechanisms and their impact further in the following Sections.

\section{Impact of Voter Registration Mechanisms}\label{s:user-parameters}

Although DAOs issue governance tokens to govern, the acquisition of these tokens does not {\em ipso facto} translate to voting power.
Users must follow some form of \stress{registration} in most DAOs to get voting rights. 
Registration is a form of \stress{admission control} for participating in governance.
Since token holders are identified only by wallet addresses on-chain, registration ensures that only ``eligible'' holders gain governance and monetary rights.
When used correctly, it can, hence, block treasury and unvested token-holding wallets (\S\ref{ss:token-types}) from participating in voting.

\begin{figure}[t]
    \centering
    \includegraphics[width=0.4\textwidth]{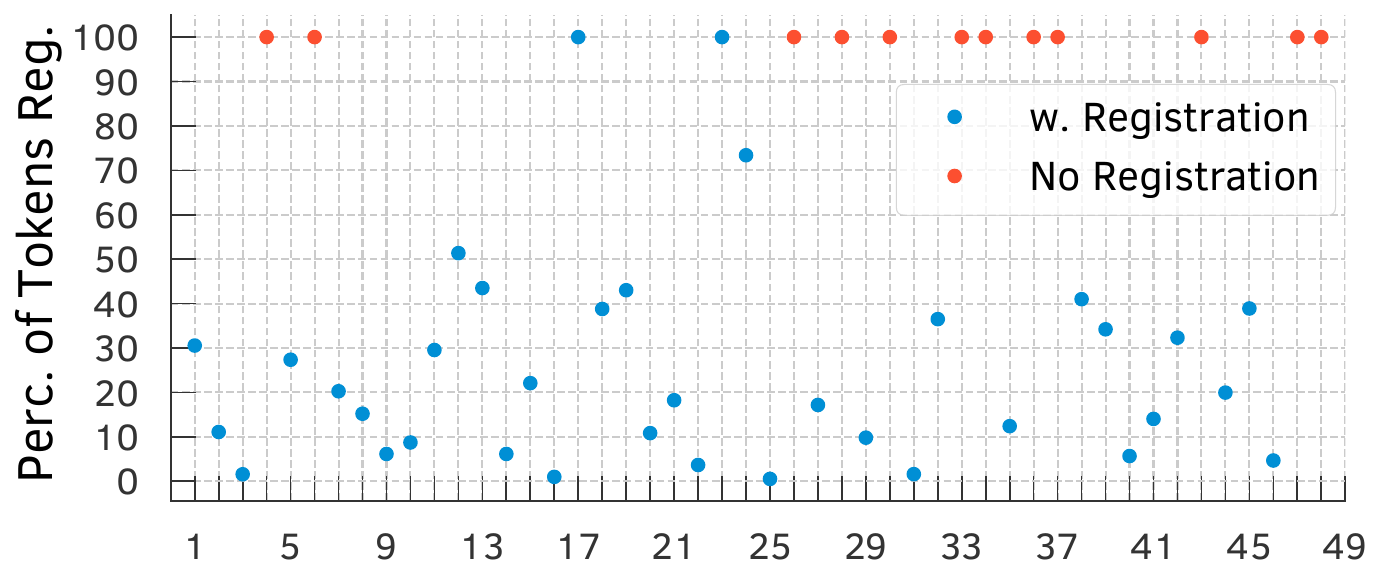}
    \figcap{The figure shows the percentage of outstanding tokens registered to vote. The x-axis values are DAO IDs from~\autoref{tab:table-registered-tokens}}
    \label{fig:perc_registered_tokens_from_outstanding}
\end{figure}

\subsection{Types of voter registration mechanisms}\label{s:voting-registration}
We classify registrations of wallet addresses into three types.

\textbf{\textit{No registration}}
is the simplest approach.
It allows \stress{any} user (or stakeholder) wallet to vote as long as they hold some specific governance tokens.
Of the $\surveysz{}$ DAOs we analyzed, 
a quarter (12) of them require no registration; they impose no admission control.

\textbf{\textit{Anonymous registration}}
\stress{requires} a user to register a wallet address in order to participate in governance, but does not request any (uniquely) identifying information about themselves, i.e. real-world identity.
It offers anonymity to token holders, but also makes it hard to reliably block wallets holding treasury and unvested tokens.
Without uniquely identifying information about users it is difficult to ascertain whether multiple wallets are controlled by the same user.
$34$ ($72\%$) DAOs in our study use anonymous registration.

\textbf{\textit{Verified registration}}, in contrast, requires a \stress{centralized} party to verify the identities of stakeholders, i.e. know-your-customer (KYC) verification, before they can participate in governance. 
The stakeholders can, hence, be held accountable for their actions; malicious users may face prosecution.
The process also enables DAOs to exclude users from certain jurisdictions, e.g., Iran and North Korea.
Only 2 DAOs---Goldfinch and Nexus Mutual---in our study use verified registration.
Goldfinch allows anybody to hold its tokens, but stipulates them to register (i.e., complete a KYC) to participate in governance.
Nexus's admission controls are stricter in contrast: They allow only the stakeholders who have completed KYC to even hold tokens.
Verified registration can reliably identify as well as filter out treasury and unvested token holders, but it also imposes a non-trivial barrier for the rest; only 0.43\% of Goldfinch's token holders are, for instance, registered.

In practice, we find that both anonymous and verified registration sharply reduce the fraction of token holders that can participate in governance votes, concentrating the voting power in few holders.
Per \autoref{fig:perc_registered_tokens_from_outstanding}, we find that only 4 out of the 36 DAOs requiring registration have more than half of their outstanding supply registered to vote, 
suggesting that most governance tokens are not actively used for governance or dividend collection.
Of these, Nouns and DxDao (\#17,23) automatically register, and Nexus Mutual requires KYC to even own tokens.
Of the 36 DAOs requiring registration, 21\% of outstanding tokens on average are registered to vote.

\begin{figure}[tb]
    \centering
    \includegraphics[width=0.4\textwidth]{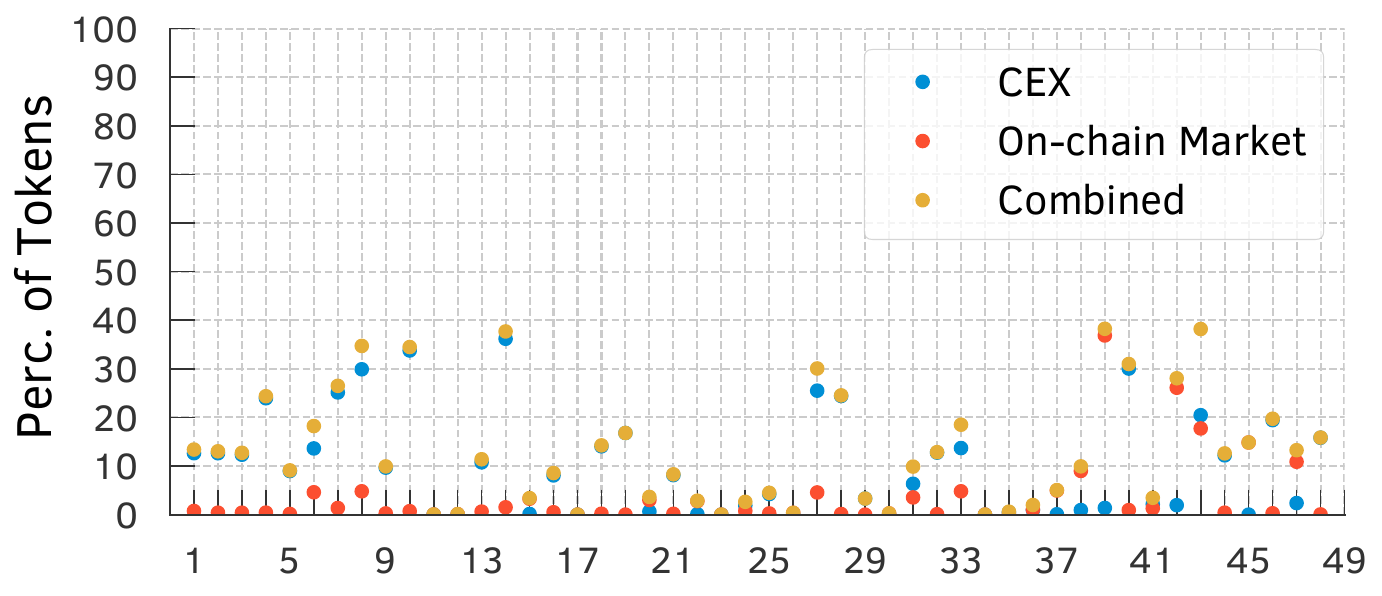}
    \figcap{We show the percentage of Outstanding Tokens in CEXes and on-chain markets. In 14 DAOs, there are more tokens in these markets than are registered to vote.}
    \label{fig:cex_dex_perc_with_outstanding_supply}
\end{figure}

\subsection{Token registration in practice}

To understand why so few eligible token holders register to vote, we investigated unregistered wallets holding large numbers of eligible tokens. 
We uncovered that many of these wallets belong to Centralized Exchanges (CEXes) or smart contracts such as lending protocols and decentralized exchanges (DEXes), where token holders transfer their tokens to earn returns for providing liquidity to these smart contracts. 
As shown in \autoref{fig:cex_dex_perc_with_outstanding_supply}, we find that on average over 10\% of outstanding tokens currently reside in CEXes, while 3.5\% reside in DEXes and lending protocols. Overall, we find that more than half (25 DAOs) have more than 10\%, and almost a quarter (11 DAOs) have more than 20\% of outstanding supply in CEXes, DEXes, and lending protocols.
Furthermore, in 14 (39\%) of the 36 DAOs requiring registration, we find that more tokens reside in these wallets than are registered to vote.\footnote{This observation establishes only a lower bound since the list of CEXes, DEXes, and lending pools is not exhaustive.}

Many users invest and hold tokens on CEXes, similar to how retail investors hold shares of public corporations via brokerages.
These tokens are held in wallets controlled by CEXes.
Similarly, token holders investing in on-chain smart contracts---such as DEXes or lending protocols---to earn returns must transfer their tokens to these contracts.
Today, these wallets are either not allowed or are being discouraged from registering to vote, disenfranchising a large fraction of eligible tokens.
For instance, many DEXes and lending protocol contracts do not offer functionality to register to vote, forcing users to forgo their governance rights, if they wish to use their tokens as collateral to earn yields. 
Similarly, in 2022, following Binance's (the world's largest CEX's) registration to vote on Uniswap (\#1~in~\autoref{tab:table-registered-tokens}), there was a huge public outcry, which eventually forced Binance to ``deregister,'' with the claim that the registration was a mistake~\cite{Binance@Coindesk,Binance-denies@coindesk}.

\subsection{On the right to register to vote}

Our observations so far raise a fundamental question about DAO governance:
{\it Should wallets of CEXes and smart contracts such as DEXes and lending protocols holding a large fraction of eligible tokens be allowed to register to vote?}
The core issue here is one of \textbf{\textit{ownership}}---are CEXes (and DEXs and lending protocols) custodians holding tokens on behalf of their owners (i.e., users of the CEXes, DEXes, and the lending protocols), or are they the actual owners of the tokens? 

The DAO community's reaction to Binance's attempt to register to vote in 2022 suggests that they do not view them as the \stress{true} owners of those tokens, but merely custodians.
Traditional finance offers, in contrast, a clear guidance---bank or brokerage deposits legally belong to depositors, who retain rights to their deposits in bankruptcy.\footnote{This is done through SIPC and FDIC insurance, which secures depositors against bankruptcy of the stockbrokers or the bank.}
However, the regulatory picture is murky and evolving when it comes to CEXes, which are legally recognized entities. 
For example, depositors to a CEX have been treated as ``unsecured creditors,'' in the bankruptcy of Celcius and FTX \cite{ftx-celcius-bankruptcy}. Depositors are considered to have essentially lent their tokens to the CEX, and the CEX essentially ``owns'' these tokens \cite{who-owns-customers-crypto@wsj}.

It is important to resolve this ambiguity around ownership to unlock the voting power of tokens held in CEXes and smart contracts.
If CEXes and smart contracts are recognized as the token owners, then their wallets should be allowed to register and exercise governance and monetary rights of the tokens.
If they are merely custodians and the owners are their users, then CEXes and smart contracts should provide new mechanisms that enable their users to exercise their right to vote.

One concern that arises with CEXes being recognized as the owners of the tokens is that it may centralize the voting power in the hands of single or a handful of CEXes.
For example, Binance has the largest holding of Lido's governance tokens. 
Since Lido (\#4~in~\autoref{tab:table-registered-tokens}) does not use registrations, Binance can not only vote on Lido's proposals but guarantee the outcomes given its stake.

{\bf Partial delegation: a mechanism to decentralize voting power of CEXes and smart contracts:}
Given our discussion, we propose that all DAOs implement and enable mechanisms such as \newterm{partial delegation}~\cite{partial-delegation@optimism}---allowing CEXes to register but delegate votes in their wallets to multiple users, according to their ownership share.
A single CEX (or DEX) wallet can register and delegate to multiple users on its platforms the monetary and governance rights attached to their tokens.
The new mechanism would mirror how shareholders holding stocks through brokerages receive dividends and, under recent regulations, can also vote, despite brokers technically holding the shares~\cite{blackrock,index-act}.
Enabling this mechanism would drastically increase the potential pool of eligible voters and bolster the decentralization of DAOs' governance.

\section{Impact of Token Staking Mechanisms}\label{s:voting-power}

When a token holder registers tokens to vote, some DAOs impose additional constraints on the tokens via a mechanism called 
\stress{token staking}.
Typically, many DAOs only require users to hold governance tokens in their wallet in order to register the tokens. 
In this approach, a user is free to sell (or buy) their tokens to (or from) another user or address at any time, but they will then lose (or gain) the governance rights associated with these tokens.
But, the ability to buy tokens to an already registered wallet and immediately obtain voting power, can be leveraged by some users to attack governance proposals.
Such a \newterm{token acquisition} attack transpired in AAVE. A single user purchased a large amount of (AAVE) tokens solely for the duration of voting of two proposals (\#27 and \#65)~\cite{dotan2023vulnerable}, whose approvals were detrimental to the protocol's stability.
Such token acquisition attacks have been observed across multiple DAOs.
A recent study of 28 real-world DAO attacks found that 10 attacks exploited the ability to buy a significant number of governance tokens on the open market, instantly gain voting rights, and steal the entirety of a DAO's treasury~\cite{feichtinger@sokattacksondao}.

\textbf{\textit{Token staking}} is a mechanism that has been proposed to prevent such attacks. 
Specifically, staking stipulates users to deposit their tokens into a specific smart contract, commonly called a “staking contract,” where the tokens are locked for a specified period; users cannot withdraw their tokens until this period has elapsed. 
This approach ties governance participation to a monetary as well as a temporal commitment. This temporal commitment is realized in two different ways.

        \point{}
        \textit{With Unlock Time}:
        Voting power here is based solely on the quantity of tokens locked.
        To withdraw their tokens back, users must initiate an unlock process and wait for a specified unlock period (typically 1-2 weeks). 
        During this unlock period, the tokens do not have any voting power.
        We observe that 6 DAOs in our study use only token staking with unlock time.
    
        \point{}
        \textit{With Lock Time}:
        Voting power, here, depends on both the quantity of tokens locked and the duration for which they are locked. The user can decide both these parameters. Voting power decays as the lock period shortens, thereby aligning users' incentives with the DAO’s long-term health. To maintain maximum influence, users must periodically re-lock their tokens for the maximum duration. We observe that 9 DAOs only use token staking with lock time.

\subsection{Token staking in practice}\label{s:token-staking-in-practice}

Token staking withstands token acquisition attacks by requiring tokens to be locked in a smart contract for a defined period.
In fact, \stress{none} of the DAOs that use token staking have been subjected to token acquisition attacks. 
An attacker risks locking tokens that could lose value following malicious governance actions, thereby rendering the attack economically infeasible.
Hence, the governance rights and the monetary rights are coupled together; governance decisions can result in loss of value in tokens, which cannot be immediately sold.
This coupling---between governance and monetary rights---in token staking, however, has lead, surprisingly, to a centralization of voting power~\cite{feichtinger@sokattacksondao,dotan2023vulnerable,lloyd2023emergent}. 
Specifically in DAOs with long lock-up periods, users may ``stake'' their tokens via an \stress{intermediary}, and we discuss next how these intermediaries subvert the resilience of the staking method.

\paraib{Staking via Intermediaries}
Angle, Balancer, Curve and Frax use token staking with the lock time model.
These four DAOs distribute dividends from profits generated by their respective DeFi protocols to the DAO stakeholders.
The dividends are proportional to voting power of the stakeholders, which depends also on the staking period (i.e., lock time) for these DAOs.
To attain maximum voting power, users must stake their tokens for the longest permissible duration (4 years for Curve, Angle, and Frax, and 1 year for Balancer) and give up their monetary right of selling the token for that duration.
To avoid this long-term locking, users often use intermediaries (e.g. Convex~\cite{convex-finance-web}, StakeDAO~\cite{stakedao-web}, and Aura~\cite{aura-web}).
These platforms stake on users’ behalf and manage the re-locking process, relieving users of the need to periodically check and renew their locks.
The intermediaries for their services take a portion of the profits collected as payment.
Most importantly, since it is the intermediaries that stake in the DAO, only the intermediaries---not the users---receive the governance rights.

When using an intermediary, users receive a 1-to-1 \textit{wrapped} version of their token, which entitles them to a share of the profits (minus the intermediary’s fee).
The wrapped tokens are freely tradable and \stress{not} subject to any lock-up period.
Users can collect the profits at any time; they, hence, relinquish their governance rights, but increase the likelihood of maximizing their economic returns.
Furthermore, the wrapped tokens are \stress{not} redeemable for the original tokens through the intermediary.
To retrieve the underlying tokens, users must trade their wrapped versions on an exchange, often at a large discount~\cite{dune@aura,dune@convex,dune@stakedao}.
Note that these wrapped tokens are similar to \stress{preferred shares} in corporate governance, which have monetary right to dividends but no governance rights.

At the time of writing, the intermediaries Convex, StakeDAO, and Aura control 53\%, 57\%, and 65\% of the voting power of Curve, Angle, and Balancer DAOs, respectively. 
Convex also holds 46\% of the voting power in Frax.
These observations strongly indicate users' preference for staking through intermediaries to derive near-maximum rewards and retain the immediate tradeability of their wrapped tokens, even if they lose their governance rights.
However, this results in massive centralization of voting power among intermediaries raising serious governance concerns for the DAO, particularly if the incentives of the DAO and the intermediary diverge.

\section{Impact of Proxy Voting Mechanisms}\label{s:voting-power-distribution}

\begin{figure*}[t]
    \centering
    \begin{subfigure}[t]{0.4\textwidth}
        \centering
        \includegraphics[width=\linewidth]{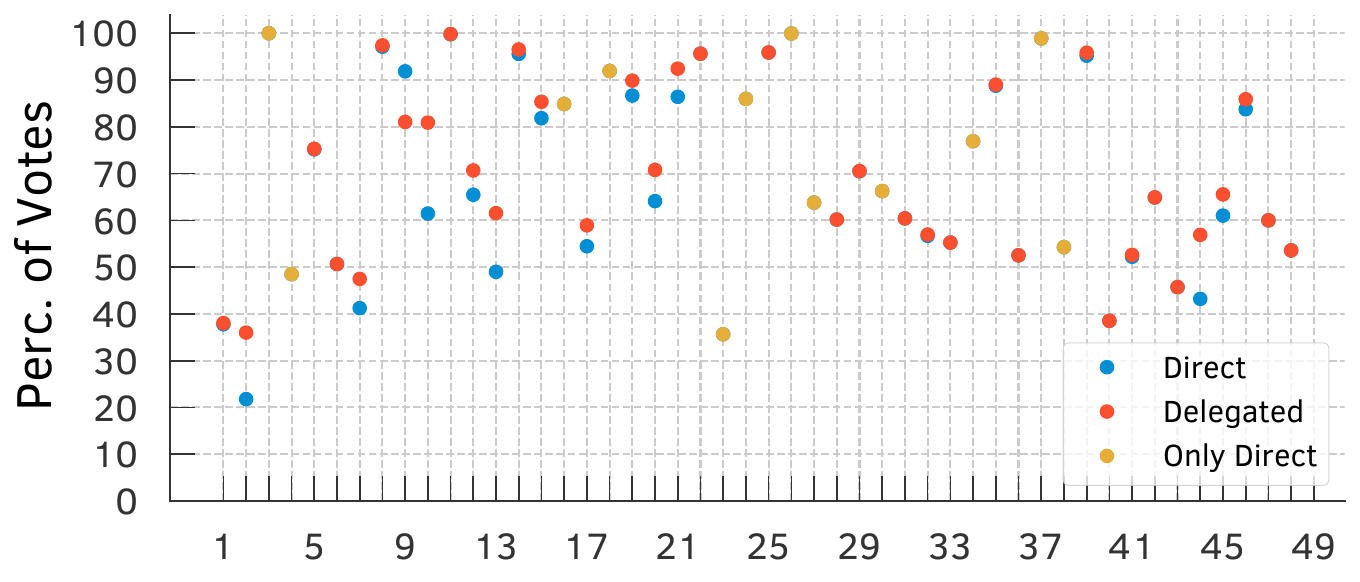}
        \sfigcap{Perc. of total Votes of top 10 voters}
        \label{fig:top_10_voters}
    \end{subfigure}
    \quad
    \begin{subfigure}[t]{0.4\textwidth}
        \centering
        \includegraphics[width=\linewidth]{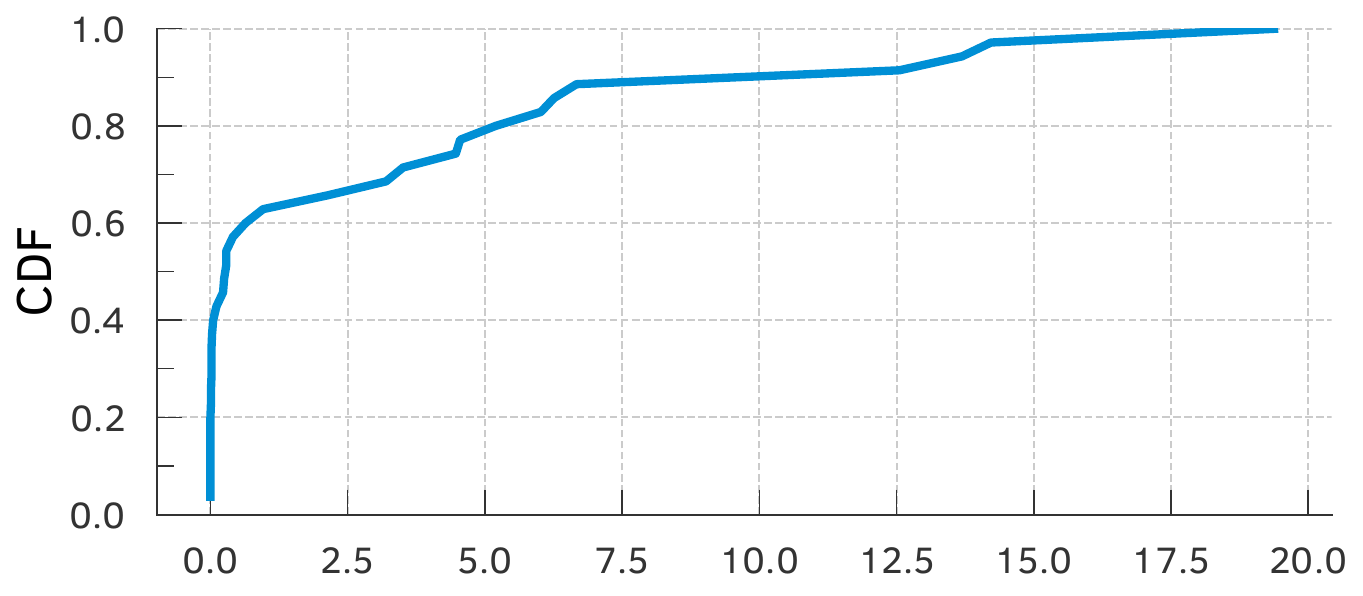}
        \sfigcap{Diff. in \% of Voting Power among top 10 Delegated and Direct voters}
        \label{fig:nakamoto_number}
    \end{subfigure}%
    \figcap{(a) 39 DAOs (81.3\%) have more than 50\% of their voting power concentrated in top 10 token holders.(b) Difference in voting power among top 10 direct and delegated wallets; delegation concentrates voting power among a few wallet addresses.}

    
    \label{fig:delegated_voting_power}
\end{figure*}

While users can register to vote, not every voter may be able to participate actively in all proposals.
Ten DAOs indeed only allow such registered wallets to participate in governance.
This restriction, however, makes them vulnerable to attacks:
Without an active participation from all of the electorate, some key proposals may fail, and attackers can exploit low participation by passing malicious proposals.
Indexed Finance, for instance, \stress{nearly} lost funds from it DAO treasury due to lack of voter engagement.
A prominent DAO member alerted the community, which resulted in many voters voting against it in the last minute~\cite{feichtinger@sokattacksondao}.
Proxy voting facilitates the formation of larger voting \stress{blocs} composed of users with smaller voting powers, and bestows power among those delegates who are better informed at making decisions on behalf of the DAO.
Hence, most DAOs in our survey allow for \stress{proxy voting} through \textit{\textbf{delegation}} and \textit{\textbf{representative voting}}.

\textbf{\textit{Delegated Voting}}\quad
In delegation, a user (the \stress{delegator}) can transfer their voting power to another user (the \stress{delegate}), effectively granting the delegate their governance right, but \stress{not} the monetary right.
If a delegate's interests are not aligned with those of a delegator, the latter can ``un-delegate'' their voting power or change the delegate at any time.
The approach essentially allows delegators to assign their voting power to delegates who are more engaged and informed about the governance of the protocol.

\textbf{\textit{Representative Voting}}\quad
Similar to delegation, in representative voting, the (voting) power to enact changes is conferred on a council of elected representatives.
The voting electorate votes for candidates in an election, and the candidates with the most votes are chosen to represent them.
Unlike delegation, however, representatives are chosen for a specific \stress{term}, and their votes cannot be simply ``un-delegated.''
Each representative has a single vote regardless of the votes they received in an election. 
In our study, only 2 DAOs---Illuvium and ParagonsDAO---use representative voting. 
Both DAOs have 7 council members serving 6-month terms.
Below, we discuss briefly how the council is formed and obtains voting power.

        \textit{Candidate Nomination}:
        Candidates, vying to become representatives, announce their candidacy on governance forums, where they state their platforms~\cite{illuvium@elections,paragonsdao@Forum}.
        Generally, there are guidelines on what information a candidate should declare, e.g., their handles on popular social media platforms such as X and discord, and questions that they should answer~\cite{paragonsdao@nominations}. 
        Anyone can self-nominate, 
        and none of the two DAOs perform \stress{any} candidate vetting.

        \textit{Election Parameters}:
        All elections are open---every voter and their choice is public, i.e. there is no secret ballot. 
        For both DAOs, the elections take place off-chain and pick the top-k candidates with the most votes. 
        Voters can divide their votes among as many candidates as they want~\cite{paragonsdao@council-voting,illuvium@elections}.

        \textit{Transfer of Power}:
        %
        %
        %
        Since elections happen off-chain in ParagonsDAO and Illuvium, power transfer from the old council 
        to the new council is \stress{not} automatic.
        The previous holders of the multi-sig wallet must remove their ownership and transfer the ownership to the newly elected council.
        There are some DAOs (not in our survey) such as Arbitrum and Synthetix, where elections happen on-chain and the transfer of power is encoded in a smart contract. 
Representative voting, however, undermines the core tenets of a DAO by adopting a structure typical of traditional organizations, where a few people manage the decision making of the organization.

\subsection{Proxy Voting in Practice}\label{s:proxy-voting-in-practice}

Delegation increases the participation in DAOs, albeit at the cost of centralizing voting powers.
Per \autoref{fig:delegated_voting_power}, delegation often centralizes voting power, since a wallet can typically delegate its entire voting power to only one other wallet.
For 20\% of DAOs in our study, the difference in the collective (or combined) voting power of the top 10 delegated voters is more than 5\% of that of the direct voters.
This voting power disparity is particularly pronounced in Gitcoin (\#10), where this difference is 19.5\%.
For about 40\% of DAOs with delegation, the difference is, however, negligible, implying that although delegation is permitted, users in these DAOs often do not delegate (and consolidate) power.
Austgen \ea argue that, \stress{in theory}, delegation can increase decentralization by redistributing the voting power of otherwise inactive users with large holdings~\cite{austgen2023dao}.
In practice, delegated voting power is, however, more centralized than direct voting power.
The only exception we observe is in the case of the 0x Protocol (\#9), which uniquely permits a single wallet to delegate its voting power across multiple wallets, thereby reducing concentration.

\paraib{Designing Decentralized Delegation}
In the current DAO landscape, CEXes cannot register or participate in governance.
Users can, moreover, delegate votes to only one address, and delegations are non-transitive (meaning if A delegates to B and B delegates to C, C does not receive A’s votes).
To enable CEXes to participate and promote greater decentralization and voter engagement, DAOs will need to support both partial and transitive delegations.
Partial delegations allows CEX wallets with large token holdings to distribute governance rights among their users, while transitive delegations would allow those users to further delegate their voting power.

\section{Related work}\label{s:related}

\paraib{Decentralized Governance}
Regular updates are essential for blockchain protocols, and agreement on such changes is often reached through informal social consensus rather than formal voting. In Bitcoin Improvement Proposals (BIPs) and Ethereum Improvement Proposals (EIPs), for instance, community discussions among users, developers, and researchers determine whether proposals gain sufficient support for implementation by core developers~\cite{bips,eips}. Deep disagreements, however, can lead to protocol splits, as observed in the forks of Bitcoin into Bitcoin and Bitcoin Cash~\cite{verge-bitcoin,coindesk-bitcoin-cash} and Ethereum into Ethereum and Ethereum Classic~\cite{kiffer2017hotnets,coindesk-ethereum-classic}. Prior work has also highlighted governance centralization in practice. Fracassi \ea show that both proposal submission and implementation within the EIP process are concentrated among a small set of actors~\cite{ethereum-centralization}. Beyond these informal processes, blockchain governance has been compared to social contract theory~\cite{reijers2016governance,chen2021decentralized} and corporate governance~\cite{allen@blockchaingov,arrunada2018blockchain,zwitter2020decentralized}, emphasizing formal voting mechanisms. 
While prior work studied different staking models~\cite{staking-paper-1,staking-paper-2,staking-paper-3}, the models do not consider unstaking time, thereby overlooking governance dynamics in DAOs with long staking periods.
Building on this literature, our study focuses on DAOs, examining how the allocation of governance and monetary rights to tokens and the design of governance mechanisms can lead to centralization.

\paraib{Drawbacks of DAO} 
Prior work of empirical studies demonstrated that voting power in DAO governance is highly centralized, calling into question the true decentralization of these protocols~\cite{fritsch@2022votingpower,messias2023gov,feichtinger2023hidden,kitzler2023governance}.
Sharma \ea analyzed the real-world operation of 10 DAOs and find limited and uneven degrees of decentralization and autonomy~\cite{sharma2023unpacking}, while Kiayias and Lazos derive seven core properties of blockchain governance and show that most of the 10 major platforms they evaluate suffer from governance deficiencies~\cite{kiayias@2022governance}.
Austgen \ea introduce a new measure of DAO centralization that highlights the effects of voter apathy, delegation, and related factors~\cite{austgen2023dao}. 
Closer to our work, Tan \ea outline open research problems across technical, economic, and legal dimensions of DAOs~\cite{tan2023open}. 
While these works identify key challenges and propose partial solutions, they do not examine the governance mechanisms currently deployed by DAOs and their implications for decentralization; in contrast, our analysis of \surveysz{} DAOs uncovers concrete design mechanisms, trade-offs, and actionable insights for improving the realization of DAOs.

\section{Concluding Remarks}\label{s:conclusion}

A DAO’s approach to assigning governance and monetary rights amongst the holders of its tokens profoundly affects its security and decentralization.
This work examines how three core mechanisms namely, voter registration, token staking, and proxy voting, for distributing these rights can inherently lead to centralization.
Unlike prior work~\cite{feichtinger@sokattacksondao,messias2023gov,sharma2023unpacking} that focus of vulnerabilities that arise from a DAO's implementation and measure voting power concentration in specific DAOs, we reveal the fundamental trade-offs within DAO design itself that drives centralization.
We hope this work paves the way for novel research and provides a solid foundation for researchers and practitioners to design and monitor DAOs.

\bibliographystyle{IEEEtran}
\bibliography{references}

\appendix

\begin{table*}[btp]
\scriptsize
\tabcap{%
An investigation of the governance structures of \surveysz{} DeFi protocols.
%
ID refers to the assigned id for the DAO, which should be used by the reader when identifying DAOs later in the paper.
`\textbf{RegReq}' refers to if voter registration is required, and it is `No Registration (\,\idToken)' or  `Anon. Registration (\,\idWallet)' or `KYC (\idIndiv)'.
`\textbf{RegMeth}' refers to Voter Registration Method takes one of $4$ values---`Token Holding (H)', `Token Depositing w/o Token Locking (D)', `Token Staking with Unlock Time (U)', `Token Staking with Lock Time (S)'
`\textbf{Voting Modality (VM)}' has three options, and they are `Direct (D)', `Delegation (\delegate{})', and `Representative (R)''.
}%
%
\begin{tabular}{@{}rcccccccccc@{}}
\toprule
\thead{ID} & \thead{Protocol} & \thead{RegReq}  &  \thead{RegMeth}   & \thead{VM} \\
\midrule

1 & \textit{Uniswap}~\cite{Governance@Uniswap} & \idWallet & H & \delegate{} \\

2 & \textit{ENS}~\cite{ens@developerdocs} & \idWallet & H & \delegate{} \\

3 & \textit{Maker}~\cite{Governance@MakerDAO} & \idWallet & D & \direct \\


4 & \textit{Lido Governance}~\cite{lido@developerdocs} & \idToken & H & \direct \\

5 & \textit{Frax Finance}~\cite{fraxfinance@developerdocs} & \idWallet & S & \delegate{} \\



6 & \textit{AAVE}~\cite{Governance@AAVE} & \idToken & H, U & \delegate{} \\

7 & \textit{Compound}~\cite{Governance@Compound} & \idWallet & H & \delegate{} \\

8 & \textit{Radicle}~\cite{radicle@developerdocs} & \idWallet & H & \delegate{} \\

9 & \textit{0x Protocol}~\cite{0x@developerdocs} & \idWallet & U & \delegate{} \\

10 & \textit{Gitcoin}~\cite{Gitcoin} & \idWallet & H & \delegate{} \\

11 & \textit{Silo Finance}~\cite{silofinance@developerdocs} & \idWallet & H & \delegate{} \\

12 & \textit{Lyra}~\cite{lyra@developerdocs} & \idWallet & U & \delegate{} \\

13 & \textit{API3}~\cite{api3@developerdocs} & \idWallet & U & \delegate{} \\

14 & \textit{Ampleforth}~\cite{ampleforth@developerdocs} & \idWallet & H & \delegate{} \\

15 & \textit{Instadapp}~\cite{instadapp@developerdocs} & \idWallet & H & \delegate{} \\

16 & \textit{Rari}~\cite{rari@developerdocs} & \idWallet & S & \direct \\

17 & \textit{NounsDAO}~\cite{nounsdao@developerdocs} & \idWallet & H & \delegate{} \\

18 & \textit{Curve}~\cite{curve@developerdocs} & \idWallet & S & \direct \\

19 & \textit{Origin}~\cite{origin@developerdocs} & \idWallet & S & \delegate{} \\

20 & \textit{Hop DAO}~\cite{hopdao@developerdocs} & \idWallet & H & \delegate{} \\

21 & \textit{Cryptex}~\cite{cryptex@developerdocs} & \idWallet & H & \delegate{} \\

22 & \textit{Angle Protocol}~\cite{angle@developerdocs} & \idWallet & S & \delegate{} \\

23 & \textit{DxDao}~\cite{dxdao@developerdocs} & \idWallet & H & \direct \\

24 & \textit{Nexus Mutual}~\cite{nexus@developerdocs} & \idIndiv & H & \direct \\

25 & \textit{Goldfinch}~\cite{goldfinch@developerdocs} & \idIndiv & H & \delegate{} \\

26 & \textit{ParagonsDAO}~\cite{paragonsdao@developerdocs} & \idToken & H, D & \representative \\

27 & \textit{Illuvium}~\cite{illuvium@developerdocs} & \idWallet & U, S & \representative \\

28 & \textit{SuperRare}~\cite{superrare@developerdocs} & \idToken & H & \delegate{} \\

29 & \textit{Mantle}~\cite{mantle@developerdocs} & \idWallet & H & \delegate{} \\

30 & \textit{Res. Hub Fdn.}~\cite{researchhub@developerdocs} & \idToken & H & \direct \\

31 & \textit{Stargate Finance}~\cite{stargate@developerdocs} & \idWallet & S & \delegate{} \\

32 & \textit{Uma}~\cite{uma} & \idWallet & U & \delegate{} \\

33 & \textit{Cowswap}~\cite{cowprotocol@developerdocs} & \idToken & H, U & \delegate{} \\

34 & \textit{Sturdy Finance}~\cite{sturdy@developerdocs} & \idToken & H & \direct \\

35 & \textit{Euler}~\cite{euler@developerdocs} & \idWallet & H & \delegate{} \\

36 & \textit{SAFE}~\cite{safe@developerdocs} & \idToken & H, U & \delegate{} \\

37 & \textit{Tokenlon}~\cite{tokenlon@developerdocs} & \idToken & H, D & \direct \\

38 & \textit{Botto}~\cite{botto@developerdocs} & \idWallet & D & \direct \\

39 & \textit{Balancer}~\cite{balancer@developerdocs} & \idWallet & S & \delegate{} \\

40 & \textit{Sushiswap}~\cite{GovernanceSushiswap} & \idWallet & D & \delegate{} \\

41 & \textit{Gearbox}~\cite{gearbox@developerdocs} & \idWallet & U, D & \delegate{} \\

42 & \textit{Paraswap}~\cite{paraswap@developerdocs} & \idWallet & U & \delegate{} \\

43 & \textit{Alchemix}~\cite{alchemix@developerdocs} & \idToken & H, D & \delegate{} \\

44 & \textit{1Inch}~\cite{1inch@developerdocs} & \idWallet & S & \delegate{} \\

45 & \textit{Shutter DAO 0x36}~\cite{shutter@developerdocs} & \idWallet & H & \delegate{} \\

46 & \textit{Yearn Finance}~\cite{yearn@developerdocs} & \idWallet & S & \delegate{} \\

47 & \textit{Shapeshift}~\cite{shapeshit@developerdocs} & \idToken & H, U & \delegate{} \\

48 & \textit{Decentraland}~\cite{decentraland@developerdocs} & \idToken & H & \delegate{} \\

\bottomrule
\end{tabular}

\label{tab:table-dao}

\end{table*}

\end{document}